\documentclass[12pt]{iopart}
\usepackage[dvips]{graphicx}
\usepackage[english]{babel}

\begin{document}

\title[Crooks' fluctuation theorem for the FLBM]%
{Crooks' fluctuation theorem for the fluctuating lattice-Boltzmann model}
\author{L Granger, M Niemann%
\footnote{Present address: Carl von Ossietzky Universit\"at Oldenburg, Institut f\"ur Physik, 26111 Oldenburg, Germany}
	and H Kantz}
\address{ Max-Planck-Institut f\"ur Physik komplexer Systeme, N\"othnitzer Stra\ss e 38, 01187 Dresden, Germany}
\ead{granger@pks.mpg.de}

\begin{abstract}
We probe the validity of Crooks' fluctuation relation on the fluctuating lattice-Boltzmann model (FLBM),
a highly simplified lattice model for a thermal ideal gas.
We drive the system between two thermodynamic equilibrium states and compute the distribution of the work performed.
By comparing the distributions of the work performed during the forward driving and time reversed driving,
   we show that the system satisfies Crooks' relation.
The results of the numerical experiment suggest that the temperature and the free energy of the system are well defined.
\end{abstract}



\section{Crooks Fluctuation Relation}
The second principle of thermodynamics states that on average one cannot extract work from a thermodynamical system coupled with only one heat bath during a cyclic process.
To illustrate this, consider a thermodynamical system that we can drive with a parameter $\lambda$,
in contact with a heat bath at constant temperature.
By switching $\lambda$ from 0 to 1, we can drive the system from one equilibrium state to another,
and by doing this, we will \emph{perform} some work $W_\textrm{\small f}$.
Consider the \emph{time reversed} or \emph{backward} experiment:
By switching $\lambda$ from 1 down to 0, we drive the system back to its initial state
\footnote{
	By state we mean \emph{macroscopic equilibrium state}
	specified by the value of $\lambda$, the temperature of the heat bath and other external constraints.
	In fact, we assume that, given a set of external constraints, the system will always relax towards
	thermodynamic equilibrium.
}
and we \emph{extract} some work $-W_\textrm{\small b}$.
The second law of thermodynamics states that on average we extract less work during the backward experiment than we perform
during the forward experiment:
\begin{equation}
\langle W_\textrm{\small f}\rangle \geq -\langle W_\textrm{\small b}\rangle.
\end{equation}
The equality in the above inequality holds if and only if the forward and the backward process 
are performed \emph{quasi-statically}, i.e., infinitely slowly such that the
system always remains in equilibrium with the heat bath.
In that case, the work $W_\textrm r$ performed during the forward process or extracted during the backward process is called the \emph{reversible work} and is
equal to the difference in free energy between the final and the initial states
\begin{equation}
W_\textrm{r} = \Delta F = F_1 - F_0.
\end{equation}
The quantity $W_\textrm d = W_\textrm{\small f} - \Delta F$ (resp. $W_\textrm d = \Delta F+W_\textrm{\small b}$) is then the \emph{dissipated work} 
during the forward (resp. backward) process.

\emph{Crooks' fluctuation relation} \cite{crooks} gives some quantitative information on the probability to dissipate a certain amount of work
during a non quasi-static process.
Consider a particular forward protocol, specified by the time dependence $\lambda_\textrm{\small f}(t)$ of the driving parameter $\lambda$.
Now also consider the \emph{time reversed} or \emph{backward} protocol obtained from the former simply by inverting the time dependence of $\lambda$:
$\lambda_\textrm{\small b}(t)=\lambda_\textrm{\small f}(t_\textrm s-t)$, where $t_\textrm s$ is the \emph{switching time}.
Crooks' fluctuation relation states that the ratio of the probability $P_\textrm{\small f}(W_\textrm d)$
of dissipating a certain amount of work $W_\textrm{d}$ using the
forward protocol
and the probability $P_\textrm{\small b}(-W_\textrm d)$
of dissipating the opposite amount $-W_\textrm{d}$ using
the time reversed protocol is given by:
\begin{equation}\label{eq:diss}
\frac{P_\textrm{\small f}(W_\textrm{d})}{P_\textrm{\small b}(-W_\textrm{d})} = \exp\left(\frac{W_\textrm{d}}{k_BT}\right),
\end{equation}
where $k_B$ is Boltzmann's constant and $T$ is the temperature of the heat bath.
Note that the dissipated work may be negative.
We expect it to be positive only on average.
Given the definition of $W_\textrm d$, the ratio on the left hand side of (\ref{eq:diss}) is the same as the ratio of the probability to perform a certain amount of work $W$
using the direct protocol over the probability to perform the opposite amount of work using the time reversed protocol.
Crooks' relation then reads:
\begin{equation}\label{eq:crooks}
\frac{P_\textrm{\small f}(W)}{P_\textrm{\small b}(-W)} = \exp\left(\frac{W-\Delta F}{k_BT}\right).
\end{equation}
A sufficient condition for the validity of this relation 
  is that the system satisfies detailed balance for fixed $\lambda$ \cite{crooks}:
\begin{equation}\label{eq:detbal}
\frac{P(A\rightarrow B)}{P(A\leftarrow B)}=\frac{\rme^{-\beta E^\lambda_B}}{\rme^{-\beta E^\lambda_A}},
\end{equation}
where $P(A\rightarrow B)$ is the probability that the system in microstate $A$ with energy $E_A^\lambda$ makes a transition to microstate $B$ with energy $E_B^\lambda$ and 
$\beta = 1/k_BT$ is the inverse temperature of the heat bath.
This condition is very restrictive and can be relaxed to the less restrictive condition of balance, see \cite{crooks2} for more details.

Since the discovery of the Jarzynski equality in 1997 \cite{jar}, fluctuation theorems have been intensely studied
experimentally \cite{cilib,exp} and theoretically \cite{stochDyn}.
The purpose of this work is to probe the validity of Crooks' fluctuation relation on a very simple, but realistic lattice gas model.

The model chosen for the numerical experiment is the fluctuating lattice-Boltzmann model (FLBM).
It constitutes one of the simplest lattice models for an ideal thermal gas.
Classical lattice-Boltzmann models are a powerful tool for the simulation of hydrodynamic systems \cite{lbm}.
However, since they usually simulate systems on macroscopic scales, they do not include thermal fluctuations.
The FLBM was developed in order to simulate the solvent in soft matter applications, where 
thermal fluctuations are not negligible \cite{dun08,dun09}.
In this work, we simulate such an ideal gas subjected to a potential field per unit mass $\phi_\lambda(\bi r)$ depending on a
certain control parameter $\lambda$, and we measure the work performed on the system by switching $\lambda$ from 0 to 1 and from
1 down to 0{} in a finite time $t_\textrm s$.

Probing Crooks' relation on the FLBM is a way to check the thermodynamic consistency of this model.
In fact, the FLBM was developed in order to correctly simulate equilibrium fluctuations,
   but has no clearly defined total energy or free energy.
By probing Crooks' relation, we mean checking that the ratio $P_\textrm{\small f}(W)/P_\textrm{\small b}(-W)$ is an exponential function of $W$ and that 
this function is independent of the specific protocol, that is the time dependence of the driving parameter $\lambda$.
We can then check that the exponent $\beta$ appearing in this exponential function is the inverse temperature $1/k_BT$ of the the heat bath.
The main result of the present paper is that this model indeed satisfies Crooks' relation \eref{eq:crooks} in a thermodynamically consistent manner.
After having briefly introduced the model, we will present the numerical experiment.
Finally, we will focus on the distribution of the work performed during the switching and show that for this particular case, the
system satisfies Crooks' fluctuation relation.
Moreover, the distribution of work turns out to be Gaussian, which permits analytical simplifications of (\ref{eq:crooks}).

\section{The fluctuating lattice-Boltzmann model}

Our purpose here is not to give a detailed description of the fluctuating-lattice Boltzmann model (FLBM).
We will simply highlight the main features of the model we are interested in and we refer the interested reader to the literature for more details.
A good introduction to lattice-Boltzmann models can be found in \cite{lbm} and the particular model used here is described in detail in \cite{dun08,dun09}.
The FLBM is a highly simplified gas model.
Its dynamics takes place in discrete time steps and on a square lattice.
At each time step, the particles move from one node of the lattice to another according to a discrete set of velocities
and then they collide with the particles sitting on the same node.
The time step and the lattice spacing are taken to be unity defining the time and length units.
The mass unit is arbitrary, but is the same throughout this document.
The mass of one particle per unit volume is denoted by $\mu$ such that if $\rho$ is the mass contained in one unit
volume, then $\rho/\mu$ is the number of particles contained in this unit volume.
Throughout this document, all the physical magnitudes are expressed in this unit system.

\begin{figure}
\centering
\includegraphics[scale=0.3]{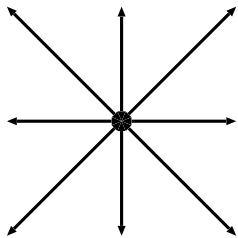}
\caption{The two dimensional nine velocities set of the D2Q9 model: 
$\bi c_0$ is the zero velocity, $\bi c_1,\dots,\bi c_4$ connect to the nearest neighbors and $\bi c_5,\dots,\bi c_8$ connect to the next nearest neighbors.
}
\label{fig:d2q9}
\end{figure}
The model used for this work is the D2Q9 model.
It consists of 9 velocities on a two dimensional square lattice, (see fig.\ref{fig:d2q9}).
Let us denote by $\{\bi c_i, i=0,\dots,8\}$ the set of velocities.
The dynamical variables of this model are the set of populations $\underline n=\{n_i, i=0,\dots,8\}$ of the different velocities.
The occupation number $n_i(\bi r,t)$ is the mass density on node $\bi r = (x,y)$ moving along velocity $\bi c_i$ at time $t$,
    where $(x,y)$ are the Cartesian components of $\bi r$.
The mass density $\rho$ and the mass current density $\bi j$ are given by the zeroth and first moments of the velocity set with respect to the set of populations:
\begin{equation}
\rho(\bi r,t) = \sum_{i=0}^8 n_i(\bi r,t),
\end{equation}
and
\begin{equation}
\bi j(\bi r,t) = \sum_{i=0}^8 n_i(\bi r,t)\bi c_i.
\end{equation}
The local velocity field $\bi v$ is then given by $\bi v(\bi r,t) = \bi j(\bi r,t)/\rho(\bi r,t)$.

The dynamics of the system takes place in two steps: propagation and collision.
During the propagation step, the particles simply move from one node to the other according to their velocity.
During the collision step, all the particles sitting on the same node
interact with one another exchanging momentum according to some basic conservation rules.
The algorithm can be summarized by the lattice-Boltzmann equation:
\begin{equation}\label{eq:lbe}
n_i(\bi r + \bi c_i,t+1) = n_i^*(\bi r,t) = n_i(\bi r,t) + \Delta_i(\underline n)+F_i.
\end{equation}
Here, $n_i^*(\bi r,t)$ is the occupation of velocity $\bi c_i$ just after the collision step and $\{\Delta_i\}$ is the collision operator.
The last term $F_i$ simulates the action of a force per unit volume $\bi F$ applied on the system.
The collision operator and the term due to the force will be presented just after a discussion of the equilibrium fluctuations.

\subsection{Equilibrium fluctuations}

As mentioned before,
in classical lattice-Boltzmann simulations, the collision operator is deterministic and simply lets the velocity  distribution $\{n_i\}$
relax towards
its local equilibrium value $\{n_i^\textrm{\small{eq}}\}$ \cite{dun09}.
The local equilibrium populations $\{n_i^\textrm{\small{eq}}\}$ are functions of the locally conserved 
quantities $\rho$ and $\bi v$ that we will specify later.
In addition to this relaxation, the FLBM allows for small fluctuations around local equilibrium.
These small fluctuations are assumed to be Gaussian in first approximation.
Taking mass and momentum conservation into account, the equilibrium probability distribution 
of the velocity occupations is \cite{dun08}:
\begin{equation}\label{eq:eqdist}
P(\{n_i\}|\rho,\bi v)\propto \exp\left(-\sum_{i=0}^8\frac{\left(n_i-n_i^\textrm{\small{eq}}\right)^2}{2\mu a^{c_i}\rho}\right)
\delta\left(\sum_{i=0}^8 n_i -\rho\right)\delta\left(\sum_{i=0}^8 n_i\bi c_i -\bi j\right),
\end{equation}
where $\delta(x)=1$ for $x=0$ and 0 otherwise.
The equilibrium fluctuations are controlled by $\mu$:
\begin{equation}\label{eq:fluct}
\langle n_i^2\rangle - \langle n_i\rangle^2 = \mu a^{c_i}\rho,
\end{equation}
The idea behind this relation is that the fluctuations in the number $n_i/\mu$ of particles having 
velocity $\bi c_i$ at a given node is proportional to the total number of particles $\rho/\mu$ on that node.
The weights $a^{c_i}$ serve to restore isotropy in the large scale limit.
They must sum up to unity, $\sum_i a^{c_i} = 1$, and must
be compatible with the symmetries of the lattice \cite{dun08,dun09} and in particular, they do not depend on the direction of the
velocity $\bi c_i$ but only on its magnitude.
For the D2Q9 lattice used here, their values are: $a^0=4/9$, $a^1=1/9$, $a^{\sqrt2}=1/36$ (see \cite{lbm}, p.69).

The particle mass per unit volume $\mu$ is our fluctuation parameter.
It is linked to the resolution of the simulation.
The limit $\mu\rightarrow 0$ corresponds to the thermodynamic limit:
decreasing $\mu$ means to describe the system on a more coarse grained scale so 
that the number of particles we describe goes to infinity and
the amplitude of the thermal fluctuations of  the
macroscopic observables decreases to zero.
On the other hand, if $\mu$ is of order one, then the fluctuations of the mass density are of the same order as the mean.

The temperature $T$ of the system is proportionnal to the fluctuation parameter $\mu$:
\begin{equation}\label{eq:sos}
k_BT=\mu c_\textrm s^2.
\end{equation}
Here, $c_\textrm s=\sqrt{\partial p/\partial \rho}$ is the \emph{isothermal speed of sound}.
\Eref{eq:sos} is a consequence of the equation of state of ideal gas
\footnote{
	Note that \eref{eq:igl} and \eref{eq:sos} are inhomogeneous.
	Indeed, $\mu$ is a mass \emph{density}: $\mu = m_p/b^2$, where $m_p$ is the mass of one particles and $b$ is the
	lattice spacing.
	But, by considering $b=1$, we write $\mu=m_p$ for simplicity of the notation.
	Moreover, we refer to $\mu$ as the \emph{fluctuation parameter} and not as the particle mass or as the
	temperature to keep in mind that it is a coarse graining parameter:
	Changing $\mu$ definitely changes the temperature of the system according to \eref{eq:sos}, but it
	also changes the number of particles through $m_p$.
	For a discussion about $\mu$, see \cite{mu}.
}
\begin{equation}\label{eq:igl}
p = \frac{\rho}{\mu}k_BT,
\end{equation}
where $p$ is the hydrodynamic pressure of the gas.
The speed of sound $c_\textrm s$ is an intrinsic quantity of the set of velocities :
it is the maximum speed at which a signal can propagate through the system.
It is of the order of $|\bi c_i|$.
For the D2Q9 model, its value in lattice units is $c_\textrm s=1/\sqrt 3$ (\cite{lbm} p.69).

\subsection{The collision operator}

The collision operator $\{\Delta_i\}$ computes the changes in the populations due to the collisions:
\begin{equation}
\Delta_i(\underline n) = n_i^*(\bi r,t) - n_i(\bi r,t),
\end{equation}
where $n_i(\bi r,t)$ is the precollisionnal value of the occupation of velocity $\bi c_i$ at node $\bi r$ and time $t$
and $n_i^*(\bi r, t)$ is its postcollisionnal value.
The role of the collision operator is to thermalize the system.
It operates as follows:
The velocity distribution $\{n_i\}$ at each node is linearly relaxed towards its local equilibrium value $\{n_i^\textrm{\small{eq}}\}$
 and this relaxation is balanced by a suitable thermal noise.
The equilibrium populations $\{n_i^\textrm{\small{eq}}\}$ depend only on the local values of $\rho$ and $\bi j$,
    and are such that mass and momentum are conserved:
\begin{equation}
\sum_{i=0}^8 n_i^\textrm{\small{eq}} = \rho
\end{equation}
\begin{equation}
\sum_{i=0}^8 n_i^\textrm{\small{eq}}\bi c_i = \bi j,
\end{equation}
and such that the equilibrium stress is correctly given by the Euler stress
\begin{equation}
\sum_{i=0}^8 n_i^\textrm{\small{eq}}c_{i\alpha}c_{i\beta} = p\delta_{\alpha\beta} + \rho v_\alpha v_\beta,
\end{equation}
where the hydrodynamic pressure is given by the ideal gas equation of state \eref{eq:igl} $p = \frac{\rho}{\mu} k_B T=\rho c_\textrm s^2$, and $\alpha$ and $\beta$ are the Cartesian coordinates.
Those conditions uniquely determine the local equilibrium populations.
They read \cite{dun08}:
\begin{equation}
n_i^\textrm{\small{eq}}(\rho,\bi v) = a^{c_i}\rho\left(1+\frac{\bi v\cdot\bi c_i}{c_\textrm s^2}+\frac{(\bi v\cdot\bi c_i)^2}{2c_\textrm s^4}
		-\frac{v^2}{2c_\textrm s^2}\right).
\end{equation}

The postcollisional state $n_i^*$ is obtained from the precollisional state $n_i$ by the collision rule
\footnote{
	This collision rule is the same as in \cite{dun08,dun09} in the case of one single relaxation time.
}
\begin{equation}\label{eq:collop}
n_i^*-n_i^{\textrm{\small{eq}}} = \gamma\left(n_i-n_i^\textrm{\small{eq}}\right) + r_i,
\end{equation}
where 
      $\gamma$ is a relaxation parameter satisfying $|\gamma|<1$ linked to the relaxation time of the velocity distribution,
      and $r_i$ is a random number corresponding to the fluctuating part of the collision process.

The conservation of mass and momentum imposes
\begin{equation}
\sum_{i=0}^8 r_i = 0,
\end{equation}
and
\begin{equation}
\sum_{i=0}^8 r_i\bi c_i = 0
\end{equation}
such that, in the end, there are only 6 independent random numbers corresponding to the 9 degrees of freedom minus 3 constraints.
The $r_i$s are drawn from a Gaussian distribution with zero mean, and
their variance is
\begin{equation}
\langle r_i^2\rangle = \mu a^{c_i}\rho\varphi,
\end{equation}
with $\varphi = \sqrt{1-\gamma^2}$ so that the variance of $n_i$ is correctly given by (\ref{eq:fluct}).
It can be shown \cite{dun08} that the collision rule \eref{eq:collop} satisfies detailed balance 
with \eref{eq:eqdist} as invariant distribution.

\subsection{Applying a body force to the system}

In order to drive the system, we apply a force $\bi f(\bi r, t)$ per unit mass on the system.
The force density applied on one lattice site is then $\bi F=\rho\bi f$.
The effect of the force is to increase the momentum density $\bi j$ by $\delta t\rho\bi F$,
   at each time step, where $\delta t=1$ is the time step.
This is done by adding a deterministic term $F_i=\rho f_i$ to the lattice-Boltzmann equation (\ref{eq:lbe}).
The $f_i$ must satisfy mass and momentum conservation, namely
\begin{equation}
\sum_{i=0}^8 f_i = 0,
\end{equation}
and
\begin{equation}
\sum_{i=0}^8 f_i\bi c_i = \bi f.
\end{equation}

As mentioned in \cite{dun09}, the definition of the local velocity $\bi v$ is no longer unique due to the discretization of the time.
Its value at each time step is included between its precollisional value $\sum_i n_i\bi c_i/\sum_i n_i$
and its post-collisional value $\sum_i n_i\bi c_i/\sum_i n_i+\bi f$.
As in \cite{dun09}, we define the local velocity as
\begin{equation}
\bi v = \frac{1}{\rho}\sum_{i=0}^8 n_i\bi c_i + \frac{1}{2}\bi f,
\end{equation}
which is the mean value between the precollisionnal and the postcollisonnal values.

In \cite{dun09}, the expression of $F_i$ consistent with hydrodynamics is derived.
Following from that, the expression for the $f_i$ is
\begin{equation}
f_i = a^{c_i}\left[\frac{\bi c_i}{c_\textrm s^2}+\frac{1+\gamma}{2}\left(\frac{\bi v\cdot\bi c_i}{c_\textrm s^4}\bi c_i-\frac{\bi v}{c_\textrm s^2}\right)
	\right]\cdot \bi f.
\end{equation}

\section{Numerical experiments and results}

We drive our system form one equilibrium state denoted by 0 to another denoted by 1
according to a particular protocol $\lambda(t)$, $\lambda(0)=0$ and $\lambda(t_\textrm s)=1$, where $t_\textrm s$ is the switching time.
The backward process is obtained by starting in the equilibrium state 1 and tuning $\lambda$ down to 0.
For each run, we record the work $W$ performed.
By repeating the forward and the backward experiments many times, we get the distribution $p_\textrm{\small f}(W)$ of the work performed during the forward process and
the distribution $p_\textrm{\small b}(-W)$ of the work extracted during the backward process.

\subsection{Direct and time-reversed protocol}
The macroscopic state of the system is controlled by applying a force field per unit mass
$\bi f_{\lambda(t)}(\bi r)=-\bi \nabla\phi_{\lambda(t)}(\bi r)$
derived from a parameter dependent potential $\phi_\lambda(\bi r)$ on the system.
The potential energy of the system is 
\begin{equation}
V_\lambda(t)=\sum_{\bi r}\rho(\bi r, t)\phi_{\lambda(t)}(\bi r).
\end{equation}
The work performed on the system during one run is
\begin{equation}\label{eq:int}
W = \int_0^{t_\textrm s}\frac{\delta W}{\delta\lambda}\frac{\mathrm d\lambda}{\mathrm dt}\mathrm dt
=\pm\int_0^1\frac{\delta W}{\delta\lambda}\mathrm d \lambda,
\end{equation}
where the plus sign holds for the forward process and the minus sign for the backward process. 
The work performed at each time step
$\delta W$ is the variation of potential energy of the system when $\lambda$ is varied by a small amount $\delta\lambda$,
while no
heat is exchanged with the heat bath:
\begin{equation}
\delta W=V_{\lambda+\delta\lambda}-V_\lambda=\sum_{\bi r}\rho(\bi r,t)\left(\phi_{\lambda+\delta\lambda}(\bi r)-\phi_\lambda(\bi r)\right).
\end{equation}

\begin{figure}
\includegraphics[scale=1.0]{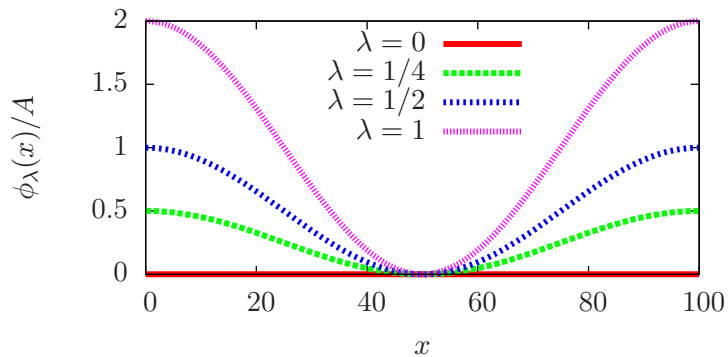}
\caption{
	Potential applied on the system as a function of $x$ expressed in units of lattice spacing.
	The forward process is started with $\lambda = 0$.
	At each time step, $\lambda$ is incremented by a small amount $\delta\lambda$ until $\lambda=1$.
}
\label{fig:pot}
\end{figure}
The potential used in this work is (see fig.\ref{fig:pot}):
\begin{equation}
\phi_\lambda(\bi r) = \lambda A \left[\cos\left(2\pi\frac{x}{l}\right)+1\right],
\end{equation}
where $l$ is the length of the system and $A$ is the amplitude of the potential.
Thus, the force applied to one unit of mass is
\begin{equation}
\bi f_\lambda = -\bi\nabla\phi_\lambda = \lambda A \frac{2\pi}{l}\sin\left( 2\pi\frac{x}{l}\right)\bi e_x.
\end{equation}
For the direct protocol $\lambda(t)=t/t_\textrm s$ and for the time-reversed protocol $\lambda(t)=1-t/t_\textrm s$.
The work $\delta W$ performed during one time step is then
\begin{equation}\label{eq:dw}
\delta W(t) = A\delta\lambda\sum_{\bi r}\rho(\bi r,t)\left[\cos\left(2\pi\frac{x}{l}\right)+1\right],
\end{equation}
where $\delta\lambda=\frac{\mathrm d\lambda}{\mathrm dt}=\pm 1/t_\textrm s$ is positive for the forward process and negative for the corresponding backward process.

The simulations were performed on a $l\times L=100\times 10$ lattice with an average density of 1000 and
periodic boundary conditions.
The relaxation parameter $\gamma$ was set to 0.9  and the amplitude $A$ of the potential to 0.01.
The parameters of the model are $\mu$, controlling the amplitude of the fluctuations, and 
the switching rate, $\delta\lambda$, controlling 
how far we are from a quasi-static process.

\subsection{Fluctuations of the work performed}

\begin{figure}
\includegraphics[scale=1.0]{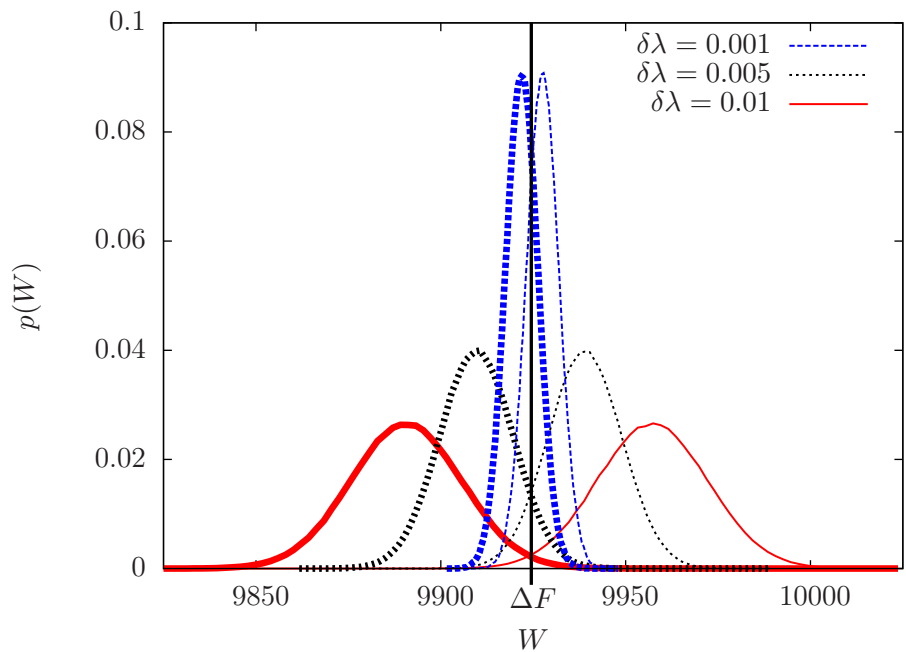}
\caption{Distribution of the work for three values of the switching rate and for $\mu=10$.
	The thin lines correspond to the work performed during the forward process 
	and the thick lines to the work extracted during the backward process.
	The work is given in lattice units.
}
\label{fig:workdist}
\end{figure}

An example of the distribution of the work performed during the forward process and the work extracted during the backward process
can be seen on figure \ref{fig:workdist}.
These simulations were performed with $\mu=10$ for various values of $\delta\lambda$.
According to Crooks' relation (\ref{eq:crooks}), the reversible work $\Delta F$ is the value of $W$ for which $p_\textrm{\small f}(W)=p_\textrm{\small b}(-W)$.
As can be seen in figure \ref{fig:workdist}, this value is independent of the switching rate $\delta\lambda$, as expected.
Moreover, we can already see that, as the switching rate decreases, the mean work converges towards 
the reversible work $\Delta F$ and the fluctuations of the work go to 0.
In the limit $\delta\lambda\rightarrow 0$, we expect the work distributions to be Dirac distributions peaked at $\Delta F$, as predicted
by equilibrium thermodynamics.

\subsubsection{Influence of the fluctuation parameter }

\begin{figure}
\includegraphics[scale=1.0]{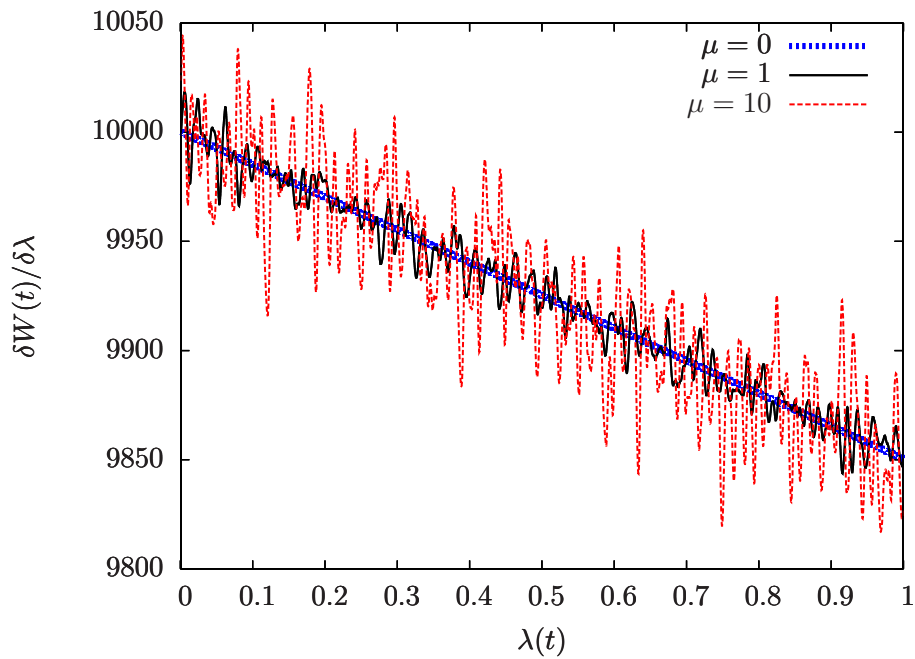}
\caption{
	Work performed at each time step as a function of $\lambda$ for three different realizations
	       of the forward process with different values of
		the fluctuation parameter $\mu$.
		Simulations performed with $\delta\lambda=10^{-4}$.
}
\label{fig:wmu}
\end{figure}

The fluctuation parameter
 $\mu$ controls the amplitude of equilibrium fluctuations of $\rho$ according to (\ref{eq:fluct}).
During the driving, the situation is not very different.
To illustrate this, we plotted examples of the evolution of the work $\delta W/\delta\lambda$ performed at every time step
during the forward process as a function of $\lambda$ for three different values of $\mu$ on figure \ref{fig:wmu}.
According to equation (\ref{eq:int}), the total work $W$ performed during the switching is the integral of this function.

\begin{figure}
\includegraphics[scale=1.0]{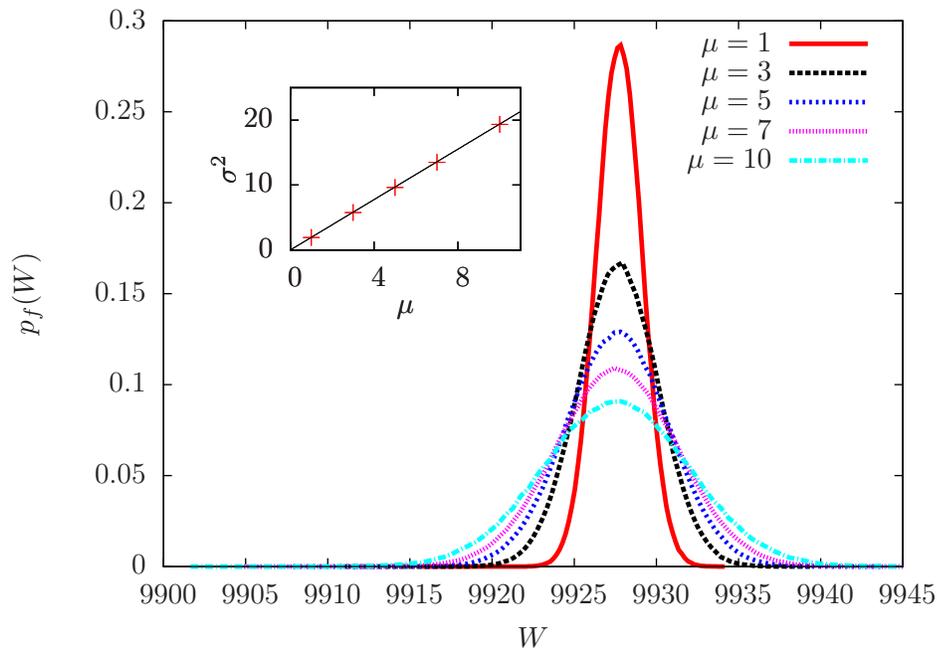}
\caption{
	Distributions of the work performed during the forward process for $\delta\lambda = 10^{-3}$ and for different values of $\mu$.
	Top left corner: variance $\sigma^2$ of these distributions as a function of $\mu$.
}
\label{fig:wdistmu}
\end{figure}

As we can see on this figure, the amplitudes of the fluctuations of $\delta W$ are clearly controlled by $\mu$.
For $\mu=0$, the evolution of the system is completely deterministic and $\delta W$ does not fluctuate,
and the bigger $\mu$ is, the larger are the fluctuations.
On figure \ref{fig:wdistmu}, we plotted the distribution $p_\textrm{\small f}(W)$ 
of the total work performed  during the forward process for various values of $\mu$.
On this figure, we can see that $\mu$ controls the fluctuations of the total work performed during the forward process 
without influencing its mean value.
In fact, the variance $\sigma^2$ of the distribution of the work is proportional to $\mu$ (see top left of fig. \ref{fig:wdistmu}).
Moreover, as we will see later, the variance of the work performed during the backward process is the same as for the corresponding
forward process.

\subsubsection{Influence of the switching rate}
\begin{figure}
\includegraphics[scale=1.0]{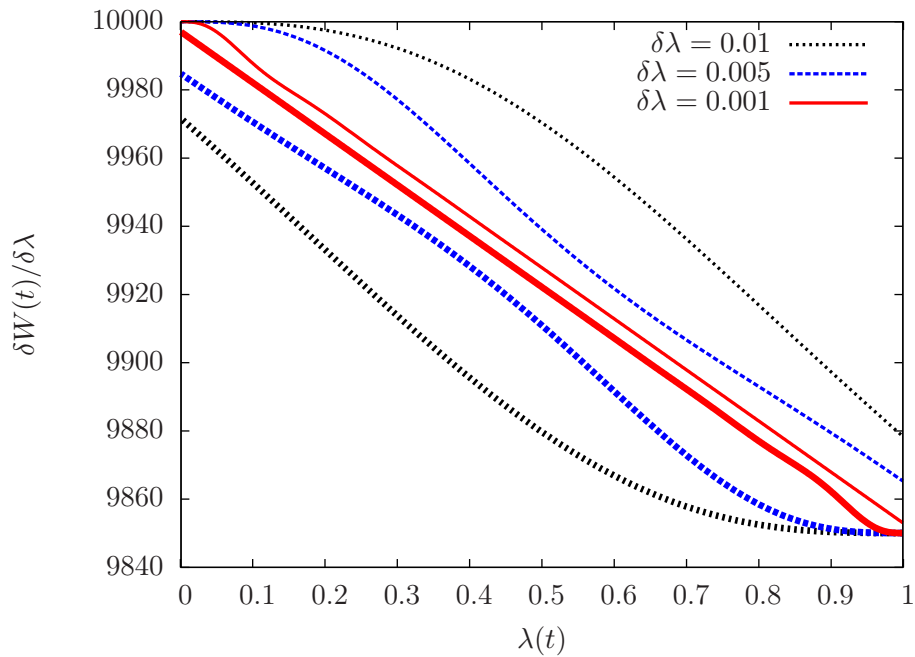}
\caption{
	Work performed during the forward process (thin lines),
	    and extracted during the backward process (thick lines) at each time step
	as a function of the driving parameter $\lambda$ in the absence of fluctuations ($\mu=0$).
}
\label{fig:nofluct}
\end{figure}
To study the influence of the switching rate on the mean work performed during the process, let us first consider the case without
fluctuations ($\mu=0$).
Figure \ref{fig:nofluct} shows $\delta W/\delta\lambda$ as a function of $\lambda$ for $\mu=0$ for various switching rates.
In that case, the work performed during the forward process is always bigger than the work extracted during the backward process.
The difference between both is linked to the work dissipated during the process and it goes to zero as the switching rate goes
to zero, that is as we approach a quasi-static process.

\begin{figure}
\includegraphics[scale=1.0]{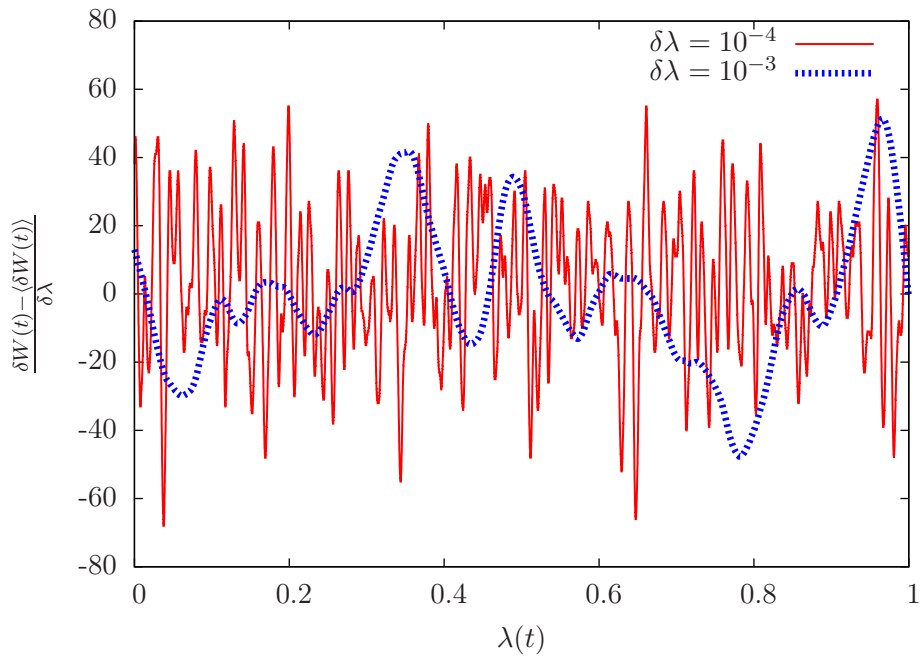}
\caption{
	Fluctuations of the
	work performed at each time step for two realizations of the experiment.
Simulations performed with $\mu$ = 10.
}
\label{fig:wlambda}
\end{figure}

For a finite amplitude of the fluctuations ($\mu>0$), the total work performed during the experiment fluctuates less for a slow switching
than for a fast switching (see fig.\ref{fig:workdist}).
However, the amplitude of the fluctuations of $\frac{\delta W}{\delta\lambda}$ are independent 
of the switching rate $\delta\lambda$.
This can be seen on figure \ref{fig:wlambda}, where we plotted the fluctuations in the work performed at every time step
$\frac{\delta W - \langle\delta W\rangle}{\delta\lambda}$ as a function of $\lambda$.
The mean work $\langle\delta W\rangle$ was obtained from a simulation with $\mu=0$.
The difference in the fluctuations of the total work $W$ comes from the fact that for a slow switching, the integral (\ref{eq:int}) is computed
over a longer time than for a fast switching, such that the fluctuations in $\frac{\delta W}{\delta\lambda}$ are partly integrated out.
For a quasi-static process, the total work $W$ does not fluctuate because the integral (\ref{eq:int}) is computed over an infinite time
such that the fluctuations in $\frac{\delta W}{\delta\lambda}$ are completely integrated out.

\subsection{Verifying Crooks' relation}

Since we perform a finite amount of realizations of the experiment,
we can probe Crooks' fluctuation theorem empirically only for values of $W$ where $p_\textrm{\small f}(W)$ and $p_\textrm{\small b}(-W)$ significantly overlap.
\begin{figure}
\includegraphics[scale=1.0]{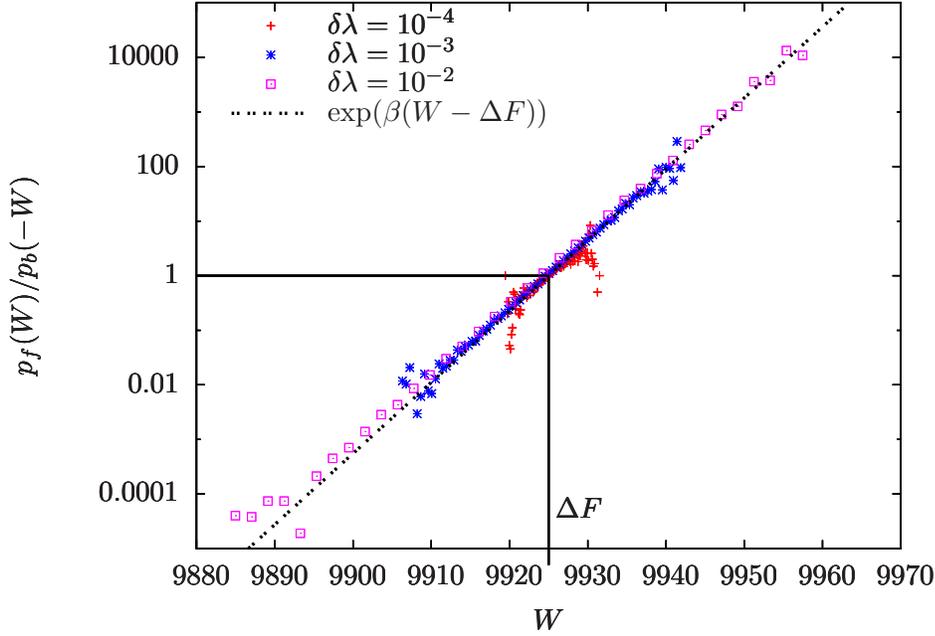}
\caption{
The ratio $p_\textrm{\small f}(W)/p_\textrm{\small b}(-W)$ as a function of $W$ for $\mu = 10$.
}
\label{fig:crooks}
\end{figure}
On figure \ref{fig:crooks}, we plotted the ratio $p_\textrm{\small f}(W)/p_\textrm{\small b}(-W)$ for $\mu=10$ for various values of the switching rate.
The line represents what we would  expect from Crooks' relation, namely $\exp\left[\beta(W-\Delta F)\right]$.
The temperature $k_BT=1/\beta=\mu c_\textrm s^2$ is given by (\ref{eq:sos}) and is 0.3{} for $\mu=10$
and the reversible work $\Delta F$ is the value of $W$ for which $p_\textrm{\small f}(W)=p_\textrm{\small b}(-W)$ and can be obtained from figure \ref{fig:workdist}.
When both $p_\textrm{\small f}(W)$ and $p_\textrm{\small b}(-W)$ are sufficiently large, their ratio satisfies Crooks' relation (\ref{eq:crooks}).

\begin{figure}
\includegraphics[scale=1.0]{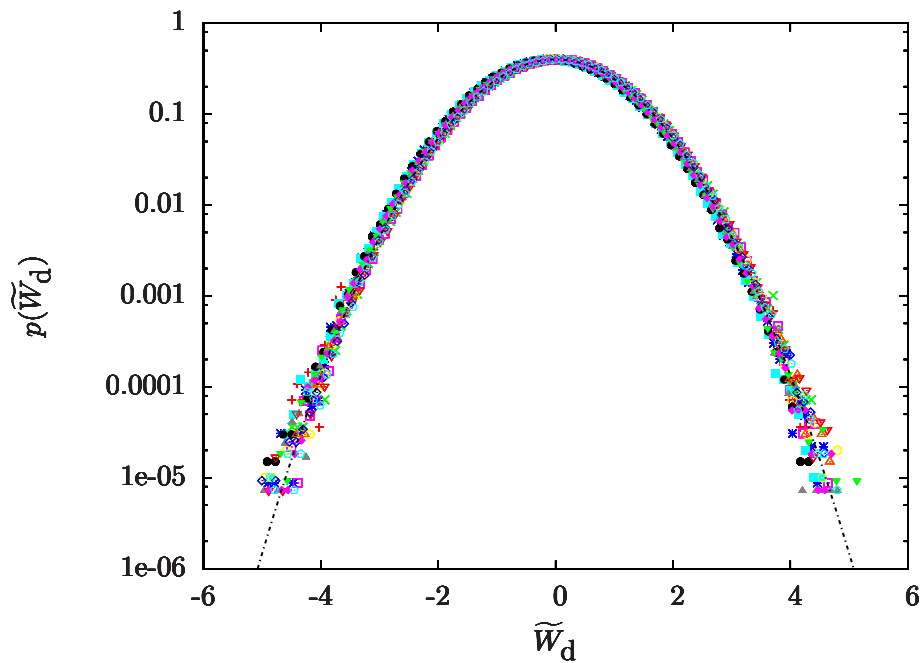}
\caption{Distribution of $\widetilde W_\textrm d = W_\textrm d/\sigma - \beta\sigma/2$ for the forward process
       and the backward process. 
The data plotted come from simulations with $\mu=10$, and $\delta\lambda = 10^{-4}$, $10^{-3}$, $5\cdot10^{-3}$, and $10^{-2}$,
    and for $\delta\lambda = 10^{-2}$ and $\mu = 3$, 5, and 7.
The line represents $1/\sqrt{2\pi}\exp(-\frac{1}{2}\widetilde W_\textrm d^2)$.
}
\label{fig:collapse}
\end{figure}

As can be seen on figure \ref{fig:collapse}, the distribution of the work is Gaussian.
As pointed out in \cite{cilib}, in that case
we can simplify Crooks' relation.
If we write the distribution of the work performed during the forward and the backward process as a Gaussian
\begin{equation}
p_\textrm {f,b}(W)=\frac{1}{\sqrt{2\pi \sigma_\textrm {f,b}^2}} \exp\left(-\frac{\left(W-\overline W_\textrm {f,b}\right)^2}{2\sigma_\textrm {f,b}^2}\right),
\end{equation}
then, the ratio appearing in Crooks' relation becomes
\begin{equation}
\fl
\frac{p_\textrm{\small f}(W)}{p_\textrm{\small b}(-W)} = \frac{\sigma_\textrm{\small b}}{\sigma_\textrm{\small f}}
\exp\left(
		\left(\frac{1}{2\sigma_\textrm{\small b}^2}-\frac{1}{2\sigma_\textrm{\small f}^2}\right)W^2
		+ \left(\frac{1}{\sigma_\textrm{\small f}^2}\overline W_f+\frac{1}{\sigma_\textrm{\small b}^2}\overline W_\textrm{\small b}\right)W
		+ \frac{\overline W_\textrm{\small b}^2}{2\sigma_\textrm{\small b}^2}-\frac{\overline W_\textrm{\small f}^2}{2\sigma_\textrm{\small f}^2}
		\right),
\end{equation}
and Crooks relation (\ref{eq:crooks}) can be valid only if the variances for the forward process and the backward process 
are equal: $\sigma_\textrm{\small f}^2=\sigma_\textrm{\small b}^2 = \sigma^2$,
    so that the term in $W^2$ vanishes in the exponent and that the factor in front of the exponential is 1.
By identification, we can express the inverse temperature $\beta$ and the reversible work $\Delta F$ in terms of the means
and the variance of the work distributions:
\begin{equation}\label{eq:beta}
\beta = \frac{\overline W_\textrm{\small f}  - \left(-\overline W_\textrm{\small b}\right)}{\sigma^2},
\end{equation}
and
\begin{equation}\label{eq:deltaF}
\Delta F = \frac{\overline W_\textrm{\small f} + \left(-\overline W_\textrm{\small b}\right)}{2}.
\end{equation}
These relations enable us to probe the validity of Crooks' relation in the Gaussian case even if the distributions for
the forward process and the backward process don't overlap.

The dissipated work ($W_{\textrm{d}}=W_\textrm{\small f}-\Delta F$ for the forward process and $W_{\textrm{d}}=W_\textrm{\small b}+\Delta F$ for the backward
		process)
is then also Gaussian, and its mean is related to its variance by
\begin{equation}\label{eq:fluctdiss}
\overline W_\textrm d = \frac{\beta\sigma^2}{2}.
\end{equation}
This relation can be seen as a generalization of the equilibrium fluctuation dissipation theorem.
The mean dissipation is proportional to the fluctuation of the dissipation and the proportionality coefficient is the temperature,
    just as in the equilibrium fluctuation dissipation theorem.

As a consequence of eq. (\ref{eq:fluctdiss}),
   the quantity 
\begin{equation}   
   \widetilde W_\textrm d = \frac{W_\textrm d-\beta\sigma^2/2}{\sigma}
\end{equation}
 is Gaussian with zero mean and unit variance,
	independently of the values of $\delta\lambda$ and $\mu$.
On figure \ref{fig:collapse}, we can see the distribution of $\widetilde W_\textrm d$ for various values of
$\mu$ and $\delta\lambda$, for the forward and for the backward process.
As we can see on this figure, the distributions of  $\widetilde W_\textrm d$ coming from simulations performed with different
values of the switching rate and of the fluctuation parameter all collapse onto a Gaussian distribution with zero mean and
unit variance
From that, we can conclude that
$p_\textrm{\small f}(W)$ and $p_\textrm{\small b}(-W)$ are Gaussian,
that the system satisfies Crooks' relation (\ref{eq:crooks}) or equivalently (\ref{eq:fluctdiss}),
and that the temperature $k_BT=1/\beta$ and the reversible work $\Delta F$ are correctly given by
Crooks' relation or equivalently by equations (\ref{eq:beta}) and (\ref{eq:deltaF}).

\section{Summary and outlook}

We have presented a basic numerical experiment which permits to probe the validity of Crooks' fluctuation relation
on the fluctuating lattice-Boltzmann model.
We have seen that this model fulfills Crooks' relation and that in this particular situation, Crooks' relation is
considerably simplified due to the Gaussian nature of the distribution of the work $W$ performed during the switching.
This simplification enables us to probe Crooks' relation, even in situations where $p_\textrm{\small f}(W)$ and $p_\textrm{\small b}(-W)$ do not overlap
significantly.

This experiment suggests that the FLBM is thermodynamically consistent.
The reversible work $\Delta F$ depends only on the initial and final states, that is on $\lambda(0)$ and $\lambda(t_\textrm s)$.
The mean dissipated work $\overline W_\textrm d$ depends only on the switching rate $\delta\lambda$.
The fluctuations $\sigma^2$ depend on the fluctuation parameter $\mu$ and on the switching rate $\delta\lambda$.
Even though this model has no a priori well defined energy or free energy,
one could define the free energy (up to an additive constant) using \eref{eq:deltaF}.
However, the link between this quantity and the microdynamics of the system is still to determine.

The fluctuating lattice-Boltzmann model is a very simple model for a thermal gas.
It enables us to study Crooks' fluctuation relation in a simple situation.
Equation (\ref{eq:int}) suggests that the interesting quantity is $\frac{\delta W}{\delta\lambda}$.
Its value during the process only depends on the instantaneous mass density distribution $\rho(\bi r, t)$.
One could, for instance, study the fluctuations in $\rho(\bi r, t)$ around its mean value during the switching
    and compare them to the equilibrium fluctuations.
More generally, this model could help identify the sources of dissipation in a finite switching-time experiment or in a non-equilibrium steady state and the link between fluctuation and dissipation in non-equilibrium situations.

\section*{References}
\bibliographystyle{unsrt}
\bibliography{biblio}{}

\end{document}